\documentclass[showpacs,preprint,amssymb]{revtex4}
\usepackage[dvips]{epsfig}
\usepackage{float}
\usepackage{graphicx}
\usepackage{dcolumn}
\usepackage{bm}
\def\be{\begin{equation}} 
\def\ee{\end{equation}}
\def\bea{\begin{eqnarray}} 
\def\eea{\end{eqnarray}}
\def\line{\hbox to \hsize}    
\def\frac #1#2{{#1\over #2}}

\def\sgn{{\rm sgn\,}}

\def\1{\mbox{\bf 1}}
\newcommand{\comment}[1]{}

\begin{document}

\title{Integer Quantum Hall Effect on a Square Lattice with Zero Net Magnetic Field}
\author{Rahul Roy}
\affiliation{University of Illinois, Department of Physics\\ 1110 W. Green St.\\
Urbana, IL 61801 USA\\E-mail: rahulroy@uiuc.edu}

\begin{abstract}
 A square lattice model which exhibits a nonzero quantized Hall conductance in a zero net magnetic field at certain values of the parameters is presented. The quantization is due to the existence of a topological winding number that characterizes the quantum Hall phases and is expected to survive the effects of weak disorder. 
\end{abstract}

\maketitle
     
       Tight binding models of electrons in a two dimensional periodic lattice in a uniform magnetic field have complex and interesting properties \cite{Azbel, Hofstadter, Wannier, Fradkin}. The spectrum in general splits into q bands when a magnetic field with a commensurate flux $ \phi= {p \over q} \phi_0 $ per plaquette is applied, and the Hall conductance of the system is given by a sum of the Chern numbers over the magnetic Brillouin zone of the filled bands \cite{TKNN}. The energy spectrum for a commensurate flux has  energy gaps which exhibit a heirarchical Hofstadter butterfly structure as $q$ is increased, and for a sequence of rational $p/q$ which approaches an irrational number, the spectrum becomes a Cantor set \cite{Hofstadter}, and self similarity is observed in the band structure \cite{Kohmoto2}.

       The integer quantum Hall transition has been studied on models with nearest and next nearest neighbour hopping terms on the square lattice \cite{Hatsugaietal,Ludwigetal}. Here we study a tight binding model on a square lattice  with nearest neighbour (NN) and next nearest neighbour (NNN) hopping terms in a {\it zero} net magnetic field. We show that it has phases with quantized Hall conductances. The phases are topologically protected against weak disorder, and transitions occur as parameters corresponding to the hopping strength and mangetic field are varied. The model also has bulk chiral fermions at certain special points in the parameter space.   
    
    Haldane had earlier proposed  a similar  model on the honeycomb lattice \cite{Haldane} to demonstrate that 
 a quantum Hall effect (which is usually associated with a strong magnetic field and degenerate Landau levels) can take place in a zero net magnetic field. The nearest-neighbour tight binding model on a honeycomb lattice has two bands and the fermi surface consists of a pair of points. Haldane showed that a periodic local magnetic field with $C_{3}$ symmetry and zero net magnetic flux per unit cell  along with an inversion symmetry breaking term can lead to a quantized Hall response for a range of strengths of the  NNN hopping term.  
  In contrast,  the NN model on the square lattice has a single band spectrum. It has, however, been known from early attempts to simulate  lattice fermions \cite{KogutSusskind} that an isotropic NN hopping model on the square lattice with $\pi $ flux per plaquette has  a pair of zeroes in its spectrum. Such a model for spinless fermions does not violate time reversal symmetry and, as a consequence of fermion doubling \cite{NielsenNinomiya}, cannot support chiral fermions. Using topological arguments, Wen and Zee \cite{WenandZee} extended the Nielsen Ninomiya theorem \cite{NielsenNinomiya} for these models by showing that, for a magnetic field  $ \phi = {p \over q} \phi_0$ per plaquette with $q$ even, the electron hopping Hamiltonian has at least $q$ families of Dirac fermions in the continuum limit. The $\pi$ flux per plaquette corresponds to $q=2$. These models for small $q$ correspond to extremely high magnetic fields that makes their experimental realization extremely difficult.
    
    While the study of the quantum Hall phases of a model with zero net magnetic flux in a square lattice is  mathematically interesting, it is probably even more difficult to observe in an experiment. However, such a model easily generalizes to a  model with spin orbit interactions where the quantum spin Hall and unconventional integer quantum Hall effect could potentially be observed. The possibility of a quantum spin Hall effect has been suggested in graphene \cite{Kaneetal1,Kaneetal2} while the ``unconventional integer quantum Hall effect" has been observed in experiment \cite{Novoselovetal, Gusyninetal}.

\begin{figure} 
  \includegraphics[scale=0.50]{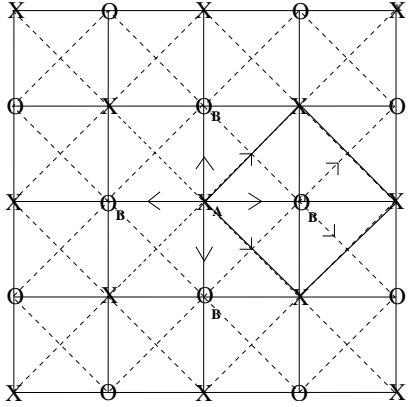}
  \caption{\label{fig1} The square lattice model showing nearest(solid lines) and next nearest neighbour hopping terms(dashed lines). The points belonging to the two sublattices A and B are marked by x's and o's. The arrows indicate the directions along which the signs accompanying the phases mentioned in the text are taken to be positive.}
  \end{figure}

     A square lattice can be thought of as consisting of two interpenetrating square sublattices, which we label A and B. Consider a tight binding model with one orbital per site on this lattice with a  nearest neighbour hopping matrix element $t$ and a next nearest neighbour hopping matrix element $t_2$, where $ t$ and $t_2$ are both real and positive (Fig.\ref{fig1}). A staggered sublattice potential $+M$ for the sites in A and $-M$ for those in B reduces the translation symmetry group of this model to that of either sublattice. The point group of the model is $C_{4v} $. Now consider the insertion of a local magnetic field $B(r)\hat{z}$  perpendicular to the plane of the lattice with  a net flux of zero  through a unit cell of either sublattice and  the translational symmetry of the sublattices.  The effect of the insertion of the magnetic field is to multiply the hopping matrix elements $t$ by  phase factors 
 $e^{i (e/\hbar) \int A.dr }$ where the integral is taken along the hopping paths which are taken to be rectilinear \cite{Haldane}. There are five independent phase factors in our problem. 
 
  We work in the gauge where the only nonzero phase for hopping elements from A to B are in the $ \pm\,\hat{x}$ directions and are both equal to  $ {\Phi \over 2}$. The phases for hopping elements from the A to A and the B to B sites in the ${\hat{x} + \hat{y}\over\sqrt{2}}$ and $ {\hat{x}-\hat{y} \over\sqrt{2}}$ directions, are labeled $\theta^A _{+},\, \theta^A _{-} $ and $ \theta^B _{+},\, \theta^B _{-} $ respectively. We restrict ourselves to studying the model, when these phases are chosen such that $ \theta^A _{+} = \theta^B_{-}= \gamma_2, \,
  \theta^A_{-} = \theta^B_{+} = \gamma_1 $.


  The Hamiltonian which acts on the momentum space  spinors  $ (\psi_A (k) ,\psi_B (k)) $ can then be written as 
  \bea
   H = 
   2 t_2 \cos(k_y a) (\cos(k_x a + {\gamma_1 + \gamma_2 \over 2}) \cos({\gamma_1 - \gamma_2 \over 2}) I + \nonumber\\
   2t(\cos(k_y a)              + 2 \cos(k_x a)\cos({\Phi\over 2}))\sigma_1 
        - 2 t\sin({\Phi \over 2})\cos(k_x a))\sigma_2 \nonumber\\ + 
       (M+2t_2  \sin(k_y a) ( \cos(k_x  a+ {\gamma_1 + \gamma_2 \over 2})\sin({\gamma_1 - \gamma_2 \over 2} )))\sigma_3
  \eea   
   where a is the length of the side of a square plaquette.

\begin{figure}
  \includegraphics[scale=0.75]{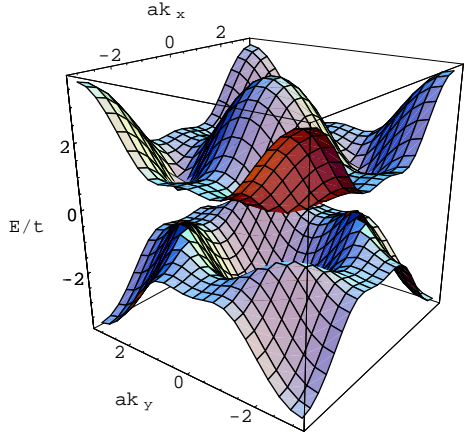}
  \caption{\label{fig2}Band structure at the point $t_2/t = 0.3, m/t=0.1,\Phi = \pi/2, \gamma_1 = 0, \gamma_2 = \pi $}
  \end{figure}

  To study the quantum Hall transition in this model, we fix the phases $ \gamma_1 = 0,\gamma_2 = {\pi} $.
   This ensures that the bands do not overlap. Then, the $T=0 $ Hall conductance of this model can be calculated using the standard Kubo  formalism, and is given by the formula as $ \sigma_{xy} = -N {e^{2} \over h}$ where N is a topological winding number which characterizes the global curvature of the momentum space berry phase \cite{Zhangetal, Stonebook}. This topological constant characterizing a two band system manifests itself in a number of problems \cite{Volovik88,Voloviketal,Krishnendu,Volovikbook}. It determines, for instance, the nature of the statistics of solitons in a p wave superfluid \cite{Voloviketal}.
    For the values of the parameters that we have chosen, we obtain 
     \bea
     \sigma_{xy} = {e^2 \over h}\sgn(\sin({\Phi\over 2}))\left\{ \matrix{ 1 & \textrm{ if $ |M| < t_2$} \cr
                              0 & \textrm {if  $|M| > t_2 $} } \right.
    \eea
     The band structure at a representative point is shown in Fig.\ref{fig2}. 
    At the points $ M = \pm t_2, \Phi = \pi $, the low energy spectrum of this system simulates relativistic chiral fermions \cite{Haldane}. The topological quantization ensures that the Hall conductance is insensitive to weak disorder.  
     
    Another region in parameter space where the integer quantum Hall transitions are observable, is when 
    $ M = 0$ and when  $ |{t_2 \over t}| < |{1 \over 2}\sin({\Phi \over 2}) \sin({\gamma_1 - \gamma_2 \over 2})| $. The latter condition ensures that the bands never overlap in this case. Then using the Kubo formalism, we obtain
    \bea
    \sigma_{xy} = {e^2 \over h}\sgn(\sin({\Phi\over 2}))\sgn(\cos(\gamma_1) - \cos(\gamma_2))
   \eea
     Here one has to vary the hopping strength $t_2 $ as one varies the phase to see the quantum Hall transition. 
     
     In the above discussion, we have examined the  quantization of the Hall conductance to non-zero values in some low dimensional manifolds in parameter space. At other points in the parameter space, there are topologically distinct phases characterized by the values of the topological constant $N$ that distinguishes them. Due to band overlap, these phases do not necessarily have a quantized Hall response. However, for minimal band overlaps, the Hall conductance is expected to remain close to its quantized value as long as the bands do not collapse. The points at which the bands touch generically have linear low energy dispersions, although the coefficients for $p_x $ and $p_y$ need not have the same magnitude. Hence these fermions will be Dirac-like fermions, rather than Dirac fermions.  
     External electric fields in the system will induce charge polarization  along the edges in an orthogonal direction.


    In Haldane's model the integer quantum Hall effect is possible with magnetic field configurations that have the symmetry group $C_3 $. These cause degeneracies at equivalent points on the Brillouin zone. A $C_{3v} $ symmetry is only possible at the zone corners when Haldane's parameters, $ M $ and $ t_2 \sin(\phi) $, are both zero, and such a symmetry causes fermion doubling, while the $ C_3 $  symmetry does not. 
     In our model, a nonzero $\Phi $ reduces the $C_{4v} $ symmetry of the original lattice to $C_{2v}$. This $C_{2v}$ symmetry, however, still restricts the values of the other phases $\theta^A _{\pm}, \,\theta^B _{\pm} $ to values at which the integer quantum Hall effect is not possible. 
    
     Our model can be extended to one which exhibits the unconventional integer quantum Hall effect and the quantum spin Hall effect with spin orbit interactions simulating magnetic fields which are opposite in direction for the two spins \cite{Rahuletal}.
   
    I would like to thank Professors Mike Stone, Eduardo Fradkin and Anthony Leggett for very useful and stimulating discussions.


\begin{thebibliography}{breitestes Label}
\bibitem{Azbel}
 M. Ya Azbel, Zh. Eksp. Teor. Fiz. {\bf 46}, 929 (1964) [Sov. Phys.—JETP {\bf 19}, 634 (1964)].
\bibitem{Hofstadter}
 D. R. Hofstadter, Phys. Rev. B {\bf 14}, 2239 (1976).
\bibitem{Wannier}
 G. H. Wannier, Phys. Status Solidi B {\bf 88}, 757 (1978) and, {\bf 93}, 337 (1979).
\bibitem{Fradkin}
Eduardo Fradkin, {\it Field Theories of Condensed Matter Systems} (Addison Wesley, Reading, MA, 1990).
\bibitem{TKNN}
D. Thouless, M. Kohmoto, M. Nightingale, and M. den Nijs, Phys. Rev. Lett. {\bf 49}, 405 (1982).
\bibitem{Kohmoto2}
 M. Kohmoto, Phys. Rev. Lett. {\bf 51}, 1198 (1983). 
 \bibitem{Hatsugaietal}
 Y. Hatsugai and M. Kohmoto, Phys. Rev. B {\bf 42}, 8282 (1990).
\bibitem{Ludwigetal}
 A. W. W. Ludwig, M. P. A. Fisher, R. Shankar, and G. Grinstein, Phys. Rev. B {\bf 50}, 7526 (1994).
\bibitem{Haldane}
F.D.M. Haldane, Phys. Rev. Lett. {\bf 61}, 2015 (1988).
\bibitem{KogutSusskind}
J. Kogut and L. Susskind, Phys. Rev. D {\bf 11}, 395 (1975).
\bibitem{NielsenNinomiya}
 H. B. Nielsen and M. Ninomiya, Nucl. Phys. B {\bf 185}, 20 (1981); B {\bf 193}, 173 (1981)
\bibitem{WenandZee}
 X. G. Wen and A. Zee, Nucl. Phys. B {\bf 316}, 641 (1989).
\bibitem{Kaneetal1}
C.L Kane, E.J.Mele, Phys. Rev. Lett {\bf 95},226801 (2005).
\bibitem{Kaneetal2}
 C.L. Kane , E.J.Mele, Phys. Rev. Lett {\bf 95}, 146802 (2005).
\bibitem{Novoselovetal}
 K.S. Novoselov, A.K. Geim, S.V. Morozov, D. Jiang, M.I. Katsnelson, I.V. Grigorieva, S.V. Dubonos, A.A. Firsov, Nature {\bf 438}, 197 (2005).
 \bibitem{Gusyninetal}
 V. P. Gusynin and S. G. Sharapov, Phys. Rev. Lett. {\bf 95}, 146801 (2005)
\bibitem{Zhangetal}
X. L. Qi, Y. S. Wu and S. C. Zhang, cond-mat/0505308 (2005).
\bibitem{Stonebook}
For a discussion of the role of the Berry phase in the usual integer quantum Hall Effect, see M. Stone, \textit{ Quantum Hall Effect} (World Scientific, Singapore, 1991). 
\bibitem{Volovik88} 
G.~E.~Volovik, Phys.\ Scr.\ {\bf 38} 321
(1988); Zh.\ Eksp.\ Teor.\  Fiz.\ {\bf 94}, 123 (1988) (English translation: Soviet Physics (JETP) {\bf 67} 1804-11 (1988)).
\bibitem{Voloviketal}
 G. E. Volovik, V. M. Yakovenko, J. Phys.  Condens.  Matter {\bf 1}, 526-5274 (1989).
\bibitem{Krishnendu}
K. Sengupta, V. M. Yakovenko, Phys. Rev. B {\bf 62}, 4586 (2000).
\bibitem{Volovikbook}
G. Volovik, The universe in a helium droplet (Oxford
Publications, Oxford, 2003).
\bibitem{Rahuletal}
Rahul Roy(in preparation). 
\end{thebibliography}
  
\end{document}